# Comment on
# "Reconsidering the nonlinear emergent inductance: time-varying Joule heating and its impact on the AC electrical response"

Tomoyuki Yokouchi[1,†], Aki Kitaori[2], Daiki Yamaguchi[2], Naoya Kanazawa[3], Max Hirschberger[1,3], Naoto Nagaosa[1], Yoshinori Tokura[1,2,4,†]

[1]*RIKEN Center for Emergent Matter Science (CEMS), Wako 351-0198, Japan.*
[2]*Department of Applied Physics, The University of Tokyo, Tokyo 113-8656, Japan*
[3]*Institute of Industrial Science, The University of Tokyo, Tokyo 153-8505, Japan*
[4]*Tokyo College, The University of Tokyo, Tokyo 113-8656, Japan.*
† To whom correspondence should be addressed.
E-mail: tomoyuki.yokouchi@riken.jp and tokura@riken.jp

## Abstract

When non-collinear spin textures are driven by current, an emergent electric field arises due to the emergent electromagnetic induction. So far, this phenomenon has been reported in several materials, manifesting the current-nonlinear imaginary part of the complex impedance. Recently, Furuta *et al.* proposed a time-varying temperature increase due to Joule heating as an alternative explanation for these current-nonlinear complex impedances [arXiv:2407.00309v1]. In this study, we re-examine the nonlinear complex impedance in $Gd_3Ru_4Al_{12}$ and $YMn_6Sn_6$, specifically addressing the impact of the time-varying temperature increase. Our findings reveal that the magnetic-field angle, frequency, and temperature dependence of nonlinear complex impedances in these two materials cannot be explained by the time-varying temperature increase. Instead, these dependencies of the imaginary part of the nonlinear impedance are consistent with the expected behaviour in the theory of emergent electromagnetic induction. Moreover, we observe a significant real part of the nonlinear complex impedance, likely resulting from the dissipation associated with the current-driven motion of helices and domain walls. Our findings highlight the diverse current-nonlinear transport phenomena of spin dynamical origin in helimagnets.

**Introduction**

The interaction between conduction electrons and spin textures gives rise to various physical phenomena. One such phenomenon is the generation of an emergent electric field due to the emergent electromagnetic induction (EEMI), when non-collinear spin textures such as helices and domain walls are driven by current. This emergent electromagnetic induction was first proposed theoretically [1]. Notably, an emergent electric field appears in the imaginary part of the complex impedance, resembling an inductance. Thus, this emergent electric field has the potential for use in novel electric devices such as nanoscale inductors and is termed "emergent inductance" [1]. Experimentally it was first reported in a helimagnet $Gd_3Ru_4Al_{12}$ [2], where various short-period non-collinear spin textures form due to the competition among the Ruderman–Kittel–Kasuya–Yosida (RKKY) interaction, magnetic anisotropy and thermal fluctuations. The variety of nearly degenerate ground states in $Gd_3Ru_4Al_{12}$ provides an ideal platform for the generation of emergent electromagnetic induction, which was experimentally found to be strongly dependent on the magnetic phases. Subsequently, EEMI is also reported in a room-temperature helimagnet $YMn_6Sn_6$ [3][4], followed by reports of EEMI from domain walls in a ferromagnet [5] and the observation of EEMI in another helimagnet, $Tb_5Sb_3$ [6].

In many experiments[2][3][4][6], the observed emergent electric fields exhibit strong nonlinearity with respect to the input current, appearing in the imaginary part of both the first- and third-harmonic complex impedance ($ImZ^{1\omega}$ and $ImZ^{3\omega}$). Additionally, the sign of the emergent electric fields can be both positive and negative, and the emergent inductance decreases in the high-frequency region (approximately 1-10 kHz). The frequency and current density dependence of the complex impedance is discussed in terms of the uniform spin canting and the phason mode, the latter of which shows the depinning transition as the current density is increased [7]. Moreover, the importance of non-adiabatic spin-transfer torque in understanding the sign of the emergent electric field has also been considered [8].

Recently, as an alternative explanation for the nonlinear complex impedances reported in these experiments, Furuta *et al.* proposed that a time-varying temperature increase can induce a current-nonlinear imaginary part of the complex impedance [9].

This model can be intuitively understood as follows: When a sine-wave current $I_0 \sin(\omega t)$ is applied to a conductor, Joule heating is proportional to the square of the sine, causing time-dependent temperature variations. This time-varying temperature variation induces the real part of the third-harmonic complex impedance ($\text{Re} Z^{3\omega}$). In particular, when the frequency is close to the thermal relaxation time, the delay in the thermal relaxation causes a delay of $\text{Re} Z^{3\omega}$, which generates the imaginary part of the first- and third-harmonic complex impedances ($\text{Im} Z^{1\omega}$ and $\text{Im} Z^{3\omega}$) proportional to the angular frequency $\omega$ up to the thermal relaxation rate.

For phenomenological calculations in the model of time-varying temperature increases[9], the first- and third-harmonic complex impedances are given by

$$\text{Re}\Delta Z^{1\omega} \equiv \text{Re}\, Z^{1\omega} - R_0(T_0) = \frac{dR_0(T_0)}{dT} \frac{P_0(I_0)}{4} [2\chi_0(T_0) + \chi'(2\omega, T_0)], \quad (1)$$

$$\text{Im}\, Z^{1\omega} = -\frac{dR_0(T_0)}{dT} \frac{P_0(I_0)}{4} \chi''(2\omega, T_0), \quad (2)$$

$$\text{Re}\, Z^{3\omega} = -\frac{dR_0(T_0)}{dT} \frac{P_0(I_0)}{4} \chi'(2\omega, T_0), \quad (3)$$

$$\text{Im}\, Z^{3\omega} = \frac{dR_0(T_0)}{dT} \frac{P_0(I_0)}{4} \chi''(2\omega, T_0), \quad (4)$$

where $T_0$, $I_0$, and $R_0$ are the temperature of heat bath, the amplitude of the AC current, and the linear resistance of the sample, respectively. $\chi^*(\omega, T_0) = \chi'(\omega, T_0) + i\chi''(\omega, T_0)$ is the thermal response function, which connects the input power $P$ to the temperature increase $\Delta T$ of the sample as follows: $\Delta T = \text{Re}(\chi^* P)$. In such models, the thermal response function is approximately described by Debye-relaxation, $\chi^*(\omega, T_0) = \frac{\chi_0(T_0)}{1+i\omega\tau_{\text{thermal}}}$, where $\chi_0(T_0)$ and $\tau_{\text{thermal}}$ are the DC limit of the thermal response function and the thermal relaxation time, respectively.

One important aspect of the time-varying temperature-increase model is the relationship among the complex impedances ($\text{Re} Z^{3\omega}$ and $\text{Im} Z^{3\omega}$) and the temperature derivative of the resistance $[dR_0(T_0)/dT]$. As seen from Eqs. (1)-(4), all the coefficients of the complex impedances include the temperature derivative of the resistance. This is because changes in the resistance caused by the temperature increases can be approximated using the first order of a Taylor expansion [9]. Additionally, $\text{Re} Z^{3\omega}$ and $\text{Im} Z^{3\omega}$ are also closely related; in particular, $-\text{Im} Z^{3\omega}/\text{Re} Z^{3\omega}$ is equal to $\chi''/\chi' = \omega \tau_{\text{thermal}}$ [see Eqs. (3) and (4)].

One other important consequence of the time-varying temperature-increase model is the relative magnitude of real and imaginary parts of the impedance in the DC and high-frequency limits; since these limits of $\chi^*(\omega, T_0)$ are given by $\lim_{\omega \to 0} \chi^*(\omega, T_0) = \chi_0(T_0)$ and $\lim_{\omega \to \infty} \chi^*(\omega, T_0) = 0$, Eqs. (1)-(4) can be described, in the DC limit, as:

$$\text{Re}\Delta Z^{1\omega} = 3 \frac{dR_0(T_0)}{dT} \frac{P_0(I_0)}{4} \chi_0(T_0), \tag{5}$$

$$\text{Re} Z^{3\omega} = 2 \frac{dR_0(T_0)}{dT} \frac{P_0(I_0)}{4} \chi_0(T_0), \tag{6}$$

$$\text{Im} Z^{1\omega} = \text{Im} Z^{3\omega} = 0, \tag{7}$$

and in the high-frequency limit as:

$$\text{Re}\Delta Z^{1\omega} = 2 \frac{dR_0(T_0)}{dT} \frac{P_0(I_0)}{4} \chi_0(T_0), \tag{8}$$

$$\text{Im} Z^{1\omega} = \text{Im} Z^{3\omega} = \text{Re} Z^{3\omega} = 0. \tag{9}$$

These two extremes, low and high frequency, can be intuitively understood as follows: In the DC limit, the time-varying temperature increase is in phase with the input current, causing a current-nonlinear resistance, which contributes to both $\text{Re} Z^{1\omega}$ and $\text{Re} Z^{3\omega}$. In contrast, in the high-frequency limit, Joule heating cannot relax within a single cycle of the input current, resulting in a constant temperature increase, which only affects $\text{Re} Z^{1\omega}$. Importantly, $\text{Re}\Delta Z^{1\omega}$ in the DC limit must always be three-halves of $\text{Re}\Delta Z^{1\omega}$ in the high-frequency limit [see Eqs. (5) and (8)].

In this study, we carefully re-examine the possibility that the nonlinear complex impedance in $Gd_3Ru_4Al_{12}$ and $YMn_6Sn_6$ would result from the time-varying temperature increase effect. We found that the nonlinear complex impedances in these materials do not exhibit the two key features of the time-varying temperature-increase model mentioned above. We also point out several qualitative aspects of the physical behaviour that are inconsistent with this model, but consistent with the EEMI scenario. Therefore, we conclude that the time-varying temperature-increase model does not explain the nonlinear complex impedances observed in these materials.

**Magnetic field angle dependence of nonlinear impedance in $Gd_3Ru_4Al_{12}$**

First, to examine whether $dR_0(T_0)/dT$ (which we write $dR/dT$ below), $ReZ^{3\omega}$, and $ImZ^{3\omega}$ satisfy Eqs. (3) and (4), we discuss the magnetic field angle dependence of $ReZ^{3\omega}$ and $ImZ^{3\omega}$ in $Gd_3Ru_4Al_{12}$. Here we used the same micro-scale devices as in Ref.[2]. The detailed experimental methods and device characterization are described in [2]. Figures 1(a-i) show the magnetic field dependence of $ReZ^{1\omega}$, $ReZ^{3\omega}$, and $ImZ^{3\omega}$ at the sample holder temperature $T_{holder}$ = 5 K measured with a frequency of $f = \omega/2\pi$ =10 kHz, a current density amplitude of $j = 1.2\times10^8$ A/m² and various angles $\theta$. The magnetic field is applied in the $ab$-plane and $\theta$ is the relative angle between the magnetic field and the current direction [see the upper panel of Fig. 1(c)]. In $Gd_3Ru_4Al_{12}$, when the magnetic field is applied in the $ab$-plane, the multi-domain proper-screw helical state (the blue

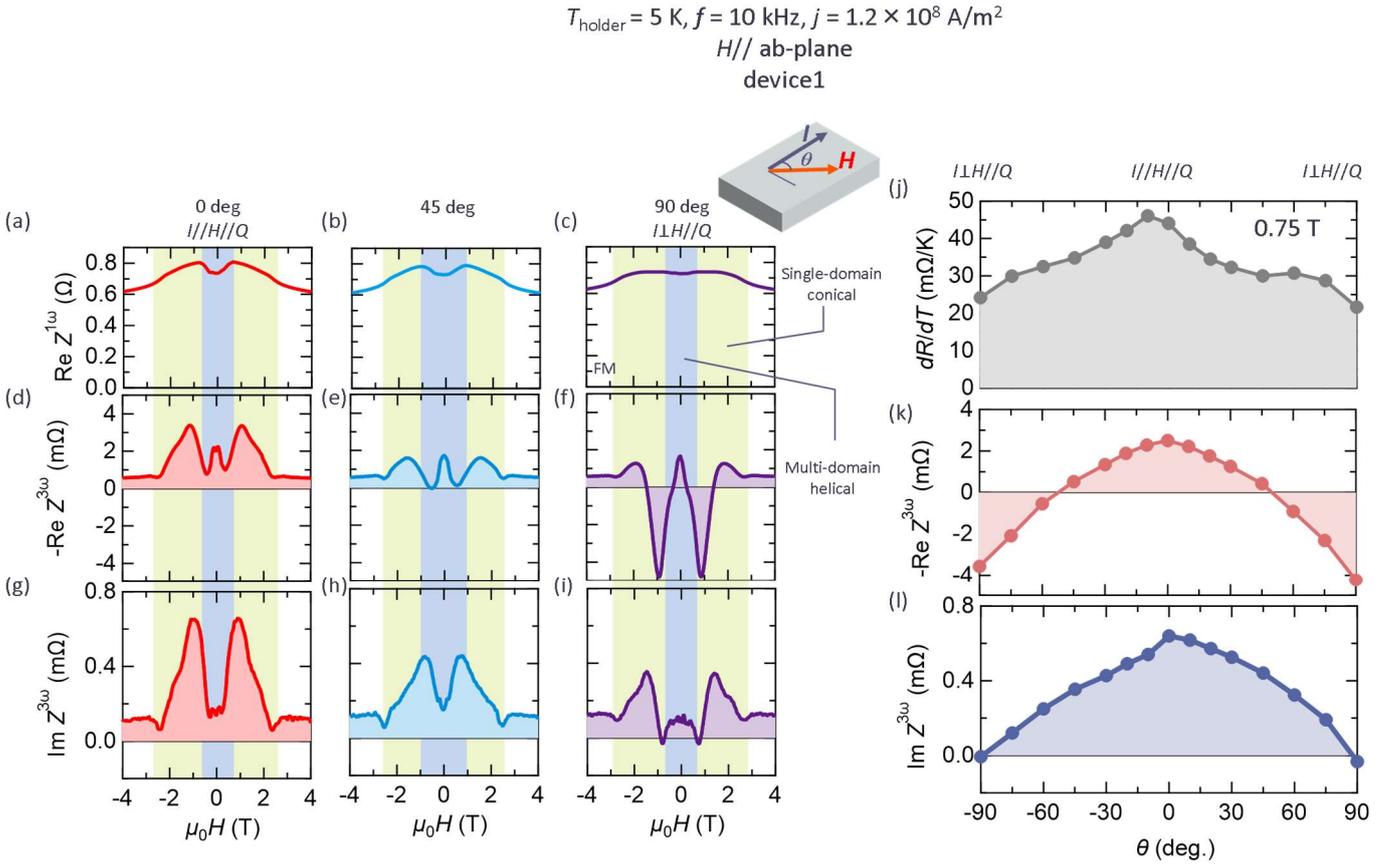

**Fig. 1** (a-i) Magnetic field ($H$) dependence of the real part of the first-harmonic complex impedance $ReZ^{1\omega}$ (a-c), the real part of the third-harmonic complex impedance -$ReZ^{3\omega}$ (d-f), and the imaginary part of the third-harmonic complex impedance $ImZ^{3\omega}$ (g-i) for various $\theta$. Here, $H$ is rotated in the $ab$-plane and $\theta$ is the relative angle between the current and $H$ as shown in the upper panel of (c). (j-l) Angle $\theta$ dependence of $dR/dT$ (j), -$ReZ^{3\omega}$ (k), and $ImZ^{3\omega}$ (l) at 0.75 T. Here, $dR/dT$ is defined as the difference between $ReZ^{1\omega}$ at 5.5 K and 5 K.

shadowed region) transforms into a single domain conical state in which the $Q$-vector is aligned parallel to $H$ (the green shadowed region) around 0.7 T [2][10]. We assigned the kinks in magnetoresistance to this multi-domain-to-single-domain transition field [Figs. 1(a-c)]. In the third-harmonic complex impedance measurement, we observe both -Re$Z^{3\omega}$ and Im$Z^{3\omega}$ [Figs. 1(d-i)]. However, importantly, the magnetic field profile of -Re$Z^{3\omega}$ is quite different from that of Im$Z^{3\omega}$. These behaviours indicate that the origins of -Re$Z^{3\omega}$ and Im$Z^{3\omega}$ are different, and Im$Z^{3\omega}$ does not result from the simple delay of -Re$Z^{3\omega}$, as proposed in the time-varying temperature-increase model. More decisively, at $\theta = 90°$, $-\text{Re}Z^{3\omega}$ exhibits a sign change around 1 T [Fig. 1(f)] while Im$Z^{3\omega}$ remains positive. As mentioned in the introduction section, $-\text{Im}Z^{3\omega}/\text{Re}\,Z^{3\omega}$ is equal to $\chi''/\chi' = \omega\tau_{\text{thermal}}$ in the time-varying temperature-increase model. Therefore, the observation of negative $-\text{Im}Z^{3\omega}/\text{Re}\,Z^{3\omega}$ implies a negative value of $\omega\tau_{\text{thermal}}$ if we assume the time-varying temperature-increase effect, which is unphysical.

We note that the magnetic field angle dependence of Im$Z^{3\omega}$ is consistent with that expected in the emergent electromagnetic induction model. Noncollinear spin textures are most efficiently driven by current via the spin-transfer torque when the current is parallel to the spin helical modulation direction, and the emergent electric field is generated parallel to the $Q$ vector in the emergent electromagnetic induction (EEMI) model [1]. Thus, Im$Z^{3\omega}$ resulting from the emergent electromagnetic induction is expected to be maximum when the current is applied parallel to $Q$ and minimum when the current is applied perpendicular to $Q$. As seen in Figs. 1(g-i), when the magnetic field is parallel to the current direction ($\theta = 0°$), Im$Z^{3\omega}$ is maximum around the multi-domain-to-single-domain transition field [Fig. 1(g)]. For $\theta = 45°$, Im$Z^{3\omega}$ is smaller than that for $Q$ // $I$. Furthermore, when $H$ is perpendicular to the current direction ($\theta = 90°$), Im$Z^{3\omega}$ becomes much smaller. These behaviours are consistent with the magnetic field angle dependence expected in the EEMI. We note that Im$Z^{3\omega}$ is not completely zero in the high-field region of the conical phase at $\theta = 90°$ and in the FM phases for all angles. These components insensitive to the magnetic field direction might result from the emergent electric fields related to the spin fluctuation as discussed in Refs. [3], [6], and [11].

For further discussion, we show the $\theta$ dependence of $dR/dT$, Re$Z^{3\omega}$, and Im$Z^{3\omega}$ in the conical phase (0.75 T) in Figs. 1(j-l). Here, we approximate $dR/dT$ as the difference between the resistance values at 5.5 K and 5 K [i.e., (Re$Z^{1\omega}$(5.5 K) - Re$Z^{1\omega}$(5.0 K))/0.5], where Re$Z^{1\omega}$ is measured with $j = 1.2\times10^8$ A/m$^2$ and $f = 10$ kHz. Again, it is confirmed that the magnetic field angle dependence of Im$Z^{3\omega}$ differs from that of -Re$Z^{3\omega}$, and the magnitude of Im$Z^{3\omega}$ is the largest for $H // I$ and the smallest for $H \perp I$. Furthermore, the angle dependence of $dR/dT$ is different from either Re$Z^{3\omega}$ or Im$Z^{3\omega}$ [Fig. 1(j)], which also indicates the origins of Re$Z^{3\omega}$ and Im$Z^{3\omega}$ are not due to the time-varying temperature increase; the Joule heating model, as seen in Eqs. (3) and (4), Re$Z^{3\omega}$ and Im$Z^{3\omega}$, would predict the proportionality to $dR/dT$. Since the thermal relaxation function [$\chi^*(\omega, T_0)$] does not depend on $\theta$, the angle dependence of $dR/dT$ should be the same as that of Re$Z^{3\omega}$ and Im$Z^{3\omega}$ in the time-varying temperature-increase model. Thus, the different magnetic field angle dependencies among $dR/dT$, Re$Z^{3\omega}$, and Im$Z^{3\omega}$ are inconsistent with the time-varying temperature-increase model proposed by Furuta et al. [9].

As short summary and relevant remark of this section, we experimentally observed both Re$Z^{3\omega}$ and Im$Z^{3\omega}$ in Gd$_3$Ru$_4$Al$_{12}$. Based on the magnetic field angle dependence, we conclude that Re$Z^{3\omega}$ and Im$Z^{3\omega}$ do not result from the time-varying temperature-increase effect. Instead, the magnetic field angle dependence of Im$Z^{3\omega}$ is consistent with the emergent electromagnetic induction due to the current-induced motion of helices. Regarding the origin of Re$Z^{3\omega}$, because the characteristic frequency of Re$Z^{3\omega}$ is similar to that of Im$Z^{3\omega}$ as presented in the next section, the origin of Re$Z^{3\omega}$ appears to be related to current-induced motion of helices, at its core, similar to Im$Z^{3\omega}$. In addition, sharp peaks of Re$Z^{3\omega}$ near the phase boundaries between multi-domain helix and single-domain conical phases imply that Re$Z^{3\omega}$ in this field region is likely related to the helical domain wall motion [Fig. 1(f)]. We assess that multiple mechanisms likely contribute to Re$Z^{3\omega}$; one major contribution to Re$Z^{3\omega}$ is the dissipation arising from current-induced helical motion and domain wall motion, while Im$Z^{3\omega}$ mainly results from the non-dissipative emergent electromagnetic induction, as driven by current-induced spin dynamics. However, a more comprehensive understanding of the origins of Re$Z^{3\omega}$ and Im$Z^{3\omega}$ is beyond the scope of this paper and remains a task for future research.

**Frequency dependence of nonlinear impedance in $Gd_3Ru_4Al_{12}$**

Next, we discuss the frequency dependence of the complex impedance in $Gd_3Ru_4Al_{12}$, which further refutes the time-varying temperature-increase effect as the origin of the observed nonlinear complex impedance. In Figs. 2(a)-(t), we show the frequency dependence of $ReZ^{1\omega}$, $Re\Delta Z^{1\omega}$, $ReZ^{3\omega}$, and $ImZ^{3\omega}$ measured in the helical phase of $Gd_3Ru_4Al_{12}$ (0 T and $T_{holder}$ = 5 K) at various current densities $j$. Here, we define $Re\Delta Z^{1\omega}$ as the change in $ReZ^{1\omega}$ from that measured at low current density ($j=0.33\times10^8$ A/m$^2$), i.e., $Re\Delta Z^{1\omega}(j) = ReZ^{1\omega}(j) - ReZ^{1\omega}(j=0.33\times10^8$ A/m$^2$) in accord to the definition employed by Furuta et al.[9]. At low current densities, $Re\Delta Z^{1\omega}$, $ReZ^{3\omega}$, and $ImZ^{3\omega}$ are almost zero [Figs. 2(f), (g), (k) and (l)]. We note that $Re\Delta Z^{1\omega}$ at $j=0.33\times10^8$ [Fig. 2(f)] is

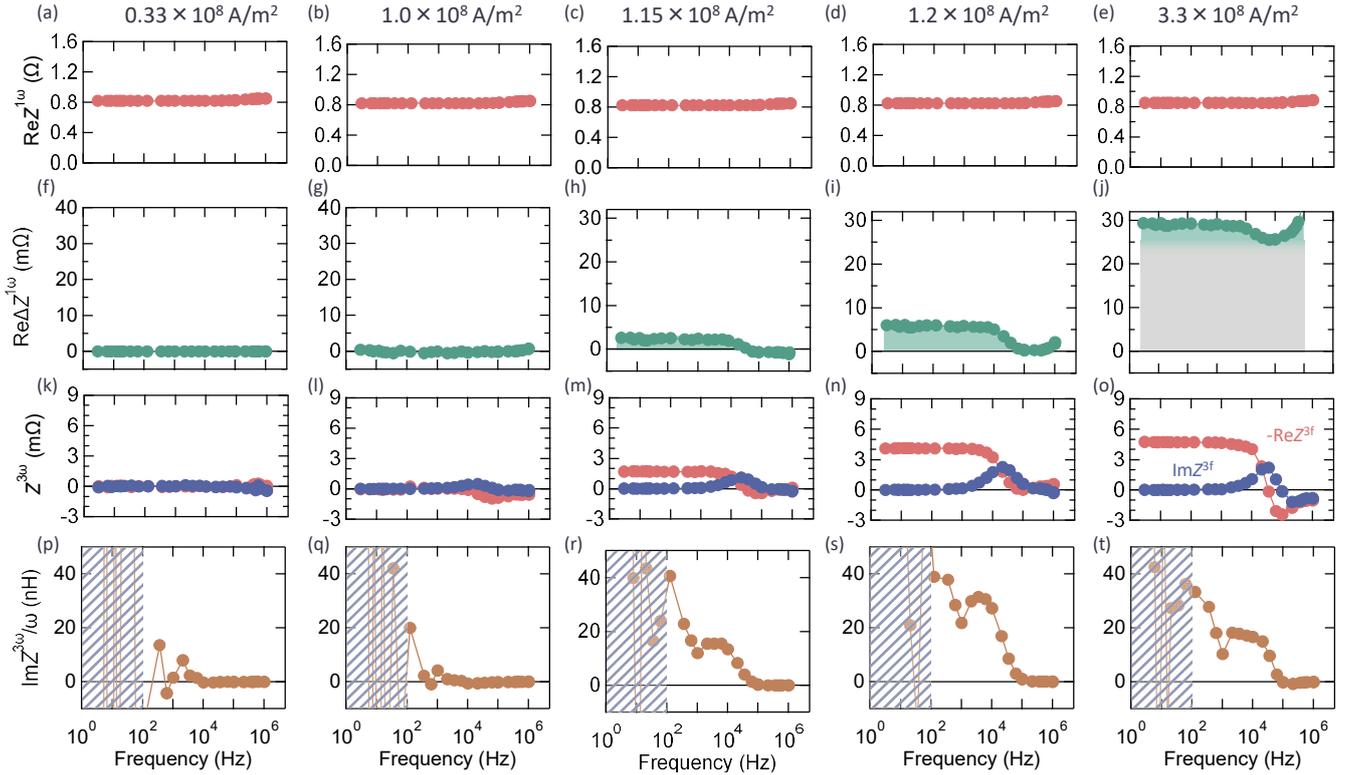

**Fig. 2** (a-e) Frequency dependence of the real part of the first-harmonic complex impedance $ReZ^{1\omega}$ for various current densities. (f-j) Frequency dependence of the change in $ReZ^{1\omega}$ compared to that measured at low current density (k-o) Frequency dependence of the real and imaginary parts of the third-harmonic complex impedance $-ReZ^{3\omega}$ and $ImZ^{3\omega}$. (p-t) Frequency dependence of the imaginary part of the third-harmonic complex impedance $ImZ^{3\omega}$ normalized by the angular frequency. In the grey-shaded region, $ImZ^{3\omega}$ was smaller than the noise, and thus we discard $ImZ^{3\omega}/\omega$ in this region.

0 by definition. In contrast, at $j$ = 1.15×10$^8$ A/m$^2$ and 1.2×10$^8$ A/m$^2$, Re$\Delta Z^{1\omega}$ is nonzero below 10 kHz and becomes zero above 100 kHz [Figs. 2(h) and (i)]. These frequency dependencies indicate the absence of the time-varying temperature increases at these current densities; if the time-varying temperature-increase model can be applied, Re$\Delta Z^{1\omega}$ is equal to $2\frac{dR_0(T_0)}{dT}\frac{P_0(I_0)}{4}\chi_0(T_0)$ in the high-frequency limit [Eq. (8)], and hence the observation of Re$\Delta Z^{1\omega}$= 0 at 100 kHz implies $\chi_0(T_0)$ = 0. Since the temperature increase $\Delta T$ is given by $\Delta T = \text{Re}[\chi^*(T_0)P] = \text{Re}\left[\frac{\chi_0(T_0)}{1+i\omega\tau_{\text{thermal}}}P\right] \propto \chi_0(T_0)$ (see the introduction for details), $\chi_0(T_0)$ = 0 means that there is no temperature increase, and thus no time-varying temperature-increase effect at $j$ = 1.15×10$^8$ A/m$^2$ and 1.2×10$^8$ A/m$^2$. Nevertheless, nonzero nonlinear complex impedance (Re$Z^{3\omega}$ and Im$Z^{3\omega}$) is clearly observed at these current densities [Figs. 2(m) and (n)], indicating that the nonlinear complex impedance in Gd$_3$Ru$_4$Al$_{12}$ does not result from the time-varying temperature-increase effect. As discussed in the previous section, Re$Z^{3\omega}$ is probably related to the dissipative nature of the current-induced dynamics of helices and domain walls. Likewise, the nonzero Re$\Delta Z^{1\omega}$ observed below 100 kHz [the green shadowed region in Figs. 2(h-i)] is also attributed to the same origin. This is because generally Re$Z^{3\omega}$ is related to a current-nonlinear Re$Z^{1\omega}$ component. The imaginary part of the complex impedance normalized by the angular frequency Im$Z^{3\omega}/\omega$ is nearly-constant values below 10 kHz, followed by a decrease above 10 kHz as shown in Fig 2(r) and (s). This tendency is consistent with the frequency dependence observed in previous reports of the emergent electromagnetic inductance of Gd$_3$Ru$_4$Al$_{12}$ [1].

At $j$ = 3.3×10$^8$ A/m$^2$, Re$\Delta Z^{1\omega}$ is finite even in the high-frequency region [Figs. 1(j)]. This additional component (grey shadowed region) is likely to result from increases in the *average* sample temperature due to Joule heating. However, the value of Re$\Delta Z^{1\omega}$ at the low-frequency region (~29 mΩ) is not three-halves of Re$\Delta Z^{1\omega}$ in the high-frequency region (~3/2×25 = 37.5 mΩ), which is a relationship that must be satisfied in the time-varying temperature-increase model. The discrepancy indicates that either Re$\Delta Z^{1\omega}$ does not reach the DC limit of the thermal relaxation even at the lowest frequency measured in the present experiment, or Re$\Delta Z^{1\omega}$ does not reach the high-frequency limit of the thermal relaxation even at the highest frequency in the experiment. In other words, the analysis implies that $\tau_{\text{thermal}}$ is outside the frequency range of our measurement. In the time-varying temperature-increase model, Im$Z^{3\omega}$ caused by the delay of Re$Z^{3\omega}$ would be appreciable near $\omega = 2\pi f \sim \tau_{\text{thermal}}^{-1}$. Therefore, we conclude that Im$Z^{3\omega}$ and Im$Z^{3\omega}/\omega$ at $j = 3.3 \times 10^8$ A/m$^2$ [Figs.2(o and t)] cannot be due to the time-varying temperature-increase effect.

**Estimation of temperature increase from transition field in Gd$_3$Ru$_4$Al$_{12}$**

To further confirm that the nonlinear impedance does not result from the time-varying temperature-increase mechanism [9], we also evaluate the temperature increase from the transverse conical-to-fan transition field ($H_{\text{TC-to-Fan}}$). As shown in Fig. 3(a), $H_{\text{TC-to-Fan}}$ in Gd$_3$Ru$_4$Al$_{12}$ strongly depends on temperature. Indeed, $H_{\text{TC-to-Fan}}$ can be well determined from kinks in the magnetoresistance [2][10]; in Fig. 3(b-c), we show the magnetic field dependence of Re$Z^{1\omega}$ measured at various current densities. The magnetic field is applied parallel to the *c*-axis. The kinks corresponding to the transverse-conical-to-fan transition are denoted by the dashed line. Apparently, $H_{\text{TC-to-Fan}}$ is robust at low current densities, but $H_{\text{TC-to-Fan}}$ decreases at high current densities. We plot $H_{\text{TC-to-Fan}}$ as a function of the current density in Fig.3 (d). Below $j = 1.7 \times 10^8$ A/m$^2$, $H_{\text{TC-to-Fan}}$ is independent of the current density, indicating that the temperature increase is negligibly small. This current density range is in accord with the range in which no heating is found (see Fig.2), based on the discussion about the frequency dependence described in the previous section. Above $1.7 \times 10^8$ A/m$^2$, $H_{\text{TC-to-Fan}}$ decreases due to the increase in the average sample temperature caused by Joule heating. The temperature increase is estimated to be 1.1 K at $j = 3.3 \times 10^8$ A/m$^2$.

In Figs. 3(e-n), we show the magnetic field dependence of Re$Z^{3\omega}$ and Im$Z^{3\omega}$ measured at various current densities. Although the Joule heating effect is negligibly small at $j = 1.2 \times 10^8$ A/m$^2$, we still observe sizable Re$Z^{3\omega}$ and Im$Z^{3\omega}$ at this current density. Again, this result supports the notion that the nonlinear complex impedance is not caused by the time-varying temperature-increase effect. We note that while an average temperature increase does occur in the high current region, this is not the cause of Im$Z^{3\omega}$ because $(\tau_{\text{thermal}})^{-1}$ is outside the frequency range of our measurement as discussed in the previous section. Additionally, the field-profiles of Re$Z^{3\omega}$ and Im$Z^{3\omega}$ change significantly depending on the current density. In particular, they do not always match the field-profile of *dR/dT* as supposed in Ref. [9]. These diverse field profiles imply that the characteristics of the current-induced dynamics of spin textures related to the origins of Re$Z^{3\omega}$ and Im$Z^{3\omega}$ such as threshold current densities for their current-driven motion are different more or less in the respective spin structures.

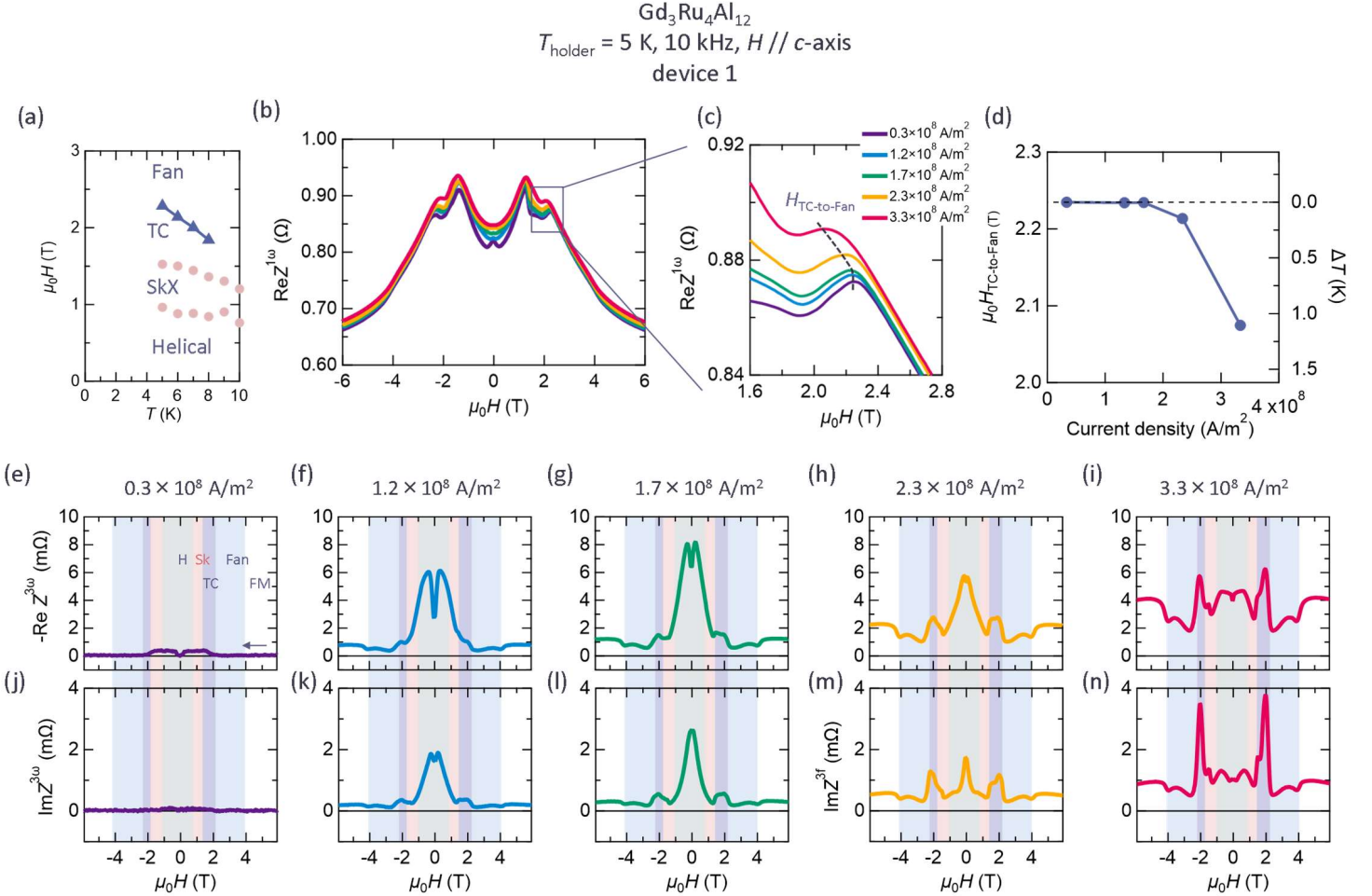

**Fig. 3** (a) Magnetic phase diagram of Gd$_3$Ru$_4$Al$_{12}$ for *H //* the *c*-axis. (b-c) Magnetic field dependence of Re$Z^{1\omega}$ measured at various current densities (b) and a magnified view (c). The dashed line in (c) indicates the transition field from the transverse conical to the fan phase ($H_{\text{TC-to-Fan}}$). (d) $H_{\text{TC-to-Fan}}$ as a function of current density. The right ordinate scale represents the estimated average temperature increase ($\Delta T$) from the base temperature (5K). (e)-(n) Magnetic-field dependence of -Re$Z^{3\omega}$ (e-i) and Im$Z^{3\omega}$ (j-n) measured at various current densities in the field-decreasing process. The green, pink, purple, light blue and white shading represents proper-screw helical (H), skyrmion (Sk), transversal conical (TC), fan and induced ferromagnetic (FM) phases, respectively. These magnetic phases are determined from measurements of magnetoresistivity and Hall conductivity ($\sigma_{xy}$) [2].

**Magnetic field dependence of Re$Z^{3\omega}$ and Im$Z^{3\omega}$ in Gd$_3$Ru$_4$Al$_{12}$**

Next, we discuss the magnetic field dependence of Re$Z^{3\omega}$ and Im$Z^{3\omega}$. Figure 4 shows the magnetic field dependence of -Re$Z^{3\omega}$, Im$Z^{3\omega}$, and $-$Im$Z^{3\omega}/$Re$Z^{3\omega}$. The magnetic field is applied parallel to the *c*-axis. As mentioned above, in the time-varying temperature-increase model, $-$Im$Z^{3\omega}/$Re$Z^{3\omega}$ is equal to $\omega\tau_{\text{thermal}}$. Here, the thermal relaxation time $\tau_{\text{thermal}}$ is determined by the extrinsic factors such as sample size and thermal contact as described in [9]. In the present case, however, $-$Im$Z^{3\omega}/$Re$Z^{3\omega}$ within the identical sample is sharply enhanced in the transverse conical phase. [Fig.4 (c)]. This correlation between $-$Im$Z^{3\omega}/$Re$Z^{3\omega}$ and the magnetic phase indicates that Re$Z^{3\omega}$ and Im$Z^{3\omega}$ are related to the spin textures, namely an intrinsic property of the material, and is inconsistent with the discussion in Ref. [9]

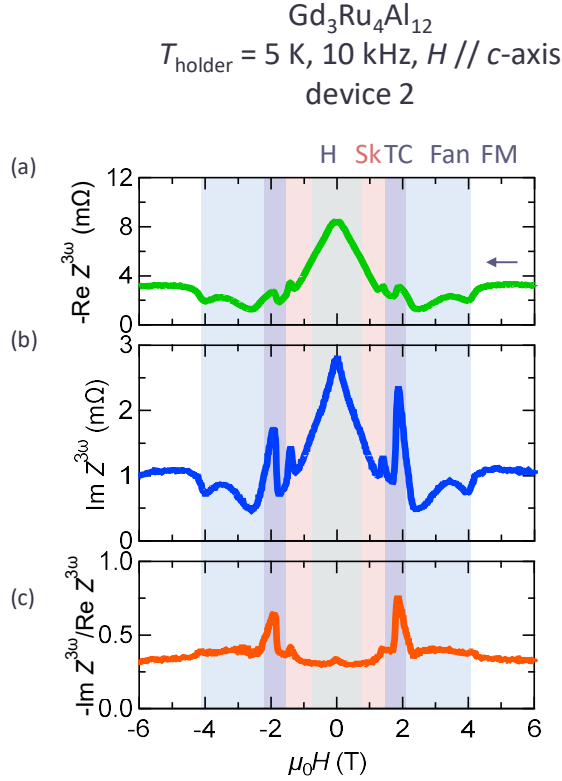

**Fig. 4** (a-c) Magnetic-field dependence of -Re$Z^{3\omega}$(a), Im$Z^{3\omega}$(b), and - Im$Z^{3\omega}/$ Re$Z^{3\omega}$(c). The green, pink, purple, light blue and white shading represents proper-screw helical (H), skyrmion (Sk), transversal conical (TC), fan and induced ferromagnetic (FM) phases, respectively.

## Temperature dependence of Im$\rho^{1f}$ and $d\rho/dT$ in YMn$_6$Sn$_6$

As the crucial test for the validity of the Joule heating effect, Furuta et al. [9] argued the similarity between the temperature dependences of Im$Z^{1\omega}$ and $dR/dT$ observed for YMn$_6$Sn$_6$ [3] on the basis of the time-varying temperature-increase model, Eq.(3). Here, we discuss in more detail the comparison of temperature dependent Im$\rho^{1f}$ and $d\rho/dT$ in YMn$_6$Sn$_6$ as well as the magnetic impurity (Tb) doped crystals whose helical magnetism is slightly modified from the parent compound YMn$_6$Sn$_6$. (Hereafter, we use the quantity of complex resistivity instead of complex impedance and frequency $f = \omega/2\pi$, following the notations of the original papers [3,4].) In Fig. 5, we show the temperature dependence of resistivity $\rho$, the imaginary part of the complex resistivity Im$\rho^{1f}$, as well as the temperature derivative of the resistivity $d\rho/dT$ in YMn$_6$Sn$_6$ and Tb-doped Y$_{1-x}$Tb$_x$Mn$_6$Sn$_6$ [3][4]. Here, the sizes of the respective devices are 9.3×37.6×4.6 μm$^3$ (width × length × thickness) for $x = 0.00$, 13.5×38.1×2.2 μm$^3$ for $x = 0.07$, and 10.1×46.8×2.9 μm$^3$ for $x = 0.10$. We show the results (Fig.5) for the same micro-scale devices as used in Refs. [3] and [4]. The detailed experimental methods and characterization of devices are described in Refs [3] and [4]. As seen in Fig. 5(d), we first

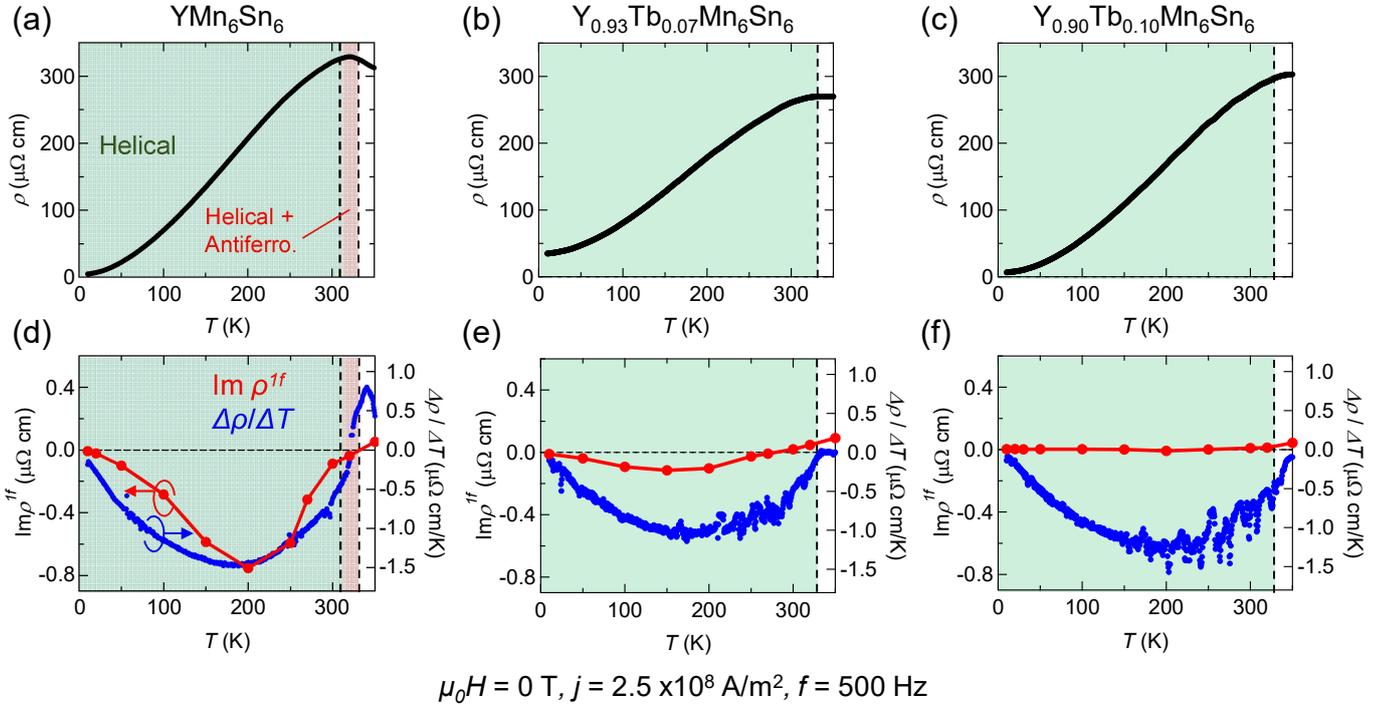

**Fig. 5** (a-f) Temperature dependence of resistivity $\rho$, the imaginary part of the complex resistivity Im$\rho^{1f}$, and the temperature derivative of the resistivity $d\rho/dT$ in YMn$_6$Sn$_6$ and Tb-doped Y$_{1-x}$Tb$_x$Mn$_6$Sn$_6$ ($x = 0.07$ and 0.10).

note that the temperature dependencies of Im$\rho^{1f}$ and $d\rho/dT$ are not clearly correlated with each other even in the YMn$_6$Sn$_6$ ($x$ = 0.00) device. In those Y$_{1-x}$Tb$_x$Mn$_6$Sn$_6$ (x = 0.00, 0.07, and 0.10) compounds, $\rho$ and $d\rho/dT$ are similar to each other in temperature dependence and of comparable magnitude except for the high temperature region above 300K where the antiferromagnetic phase coexistent with the helix phase for $x$=0.00 disappears for $x$= 0.07 and 0.10. Contrary to this moderate effect on the transport behaviour (below 300K), the impact of of Tb doping on Im$\rho^{1f}$ is huge as reported in Ref. [4]; Im$\rho^{1f}$ becomes much suppressed in magnitude at $x$ = 0.07 and is almost zero at $x$ = 0.10 over the whole temperature region as compared with the $x$=0.00 sample. We note that this drastic suppression of Im$\rho^{1f}$ in Tb-doped Y$_{1-x}$Tb$_x$Mn$_6$Sn$_6$ is due to the pinning effect of magnetic impurities (Tb) on the current-induced dynamics of the helices, as already discussed in Ref. [4]. Considering the similar size and shape of the FIB-fabricated devices, these uncorrelated behaviours between magnitudes of $d\rho/dT$ and Im$\rho^{1f}$ for a wide temperature region in the respective Tb-undoped/doped compounds are inconsistent with the time-varying temperature-increase model, even if possible slight modulation in heat capacity and thermal conductivity by Tb doping are taken into account. In other words, the Im$\rho^{1f}$ in the present case also results from the current induced dynamics of the helices which is sensitive to magnetic disorder or related pinning effect, and not from the magnitude of $d\rho/dT$.

**Conclusion**

   We carefully examined the possible current-induced Joule heating effect on the nonlinear complex impedance in Gd$_3$Ru$_4$Al$_{12}$ and YMn$_6$Sn$_6$ in response to the critical comment by Furuta et al. [9] that the nonlinear complex impedance results from the time-varying temperature-increase (TVTI) effect, not from the emergent electromagnetic induction (EEMI) effects reported in Refs. [2][3][4]. For Gd$_3$Ru$_4$Al$_{12}$, the magnetic field angle dependence and the frequency dependence of the nonlinear complex impidance are shown to be inconsistent with the TVTI model. For YMn$_6$Sn$_6$, the large variation of magnitude of the complex impedance with magnetic ion (Tb) doping while keeping the temperature-dependent resistivity similar does not match the prediction of TVTI model.

Therefore, we conclude that the TVTI model cannot explain the nonlinear complex impedance observed in these materials. Instead, the observed behaviour of the imaginary part of the nonlinear impedance is consistent with the expected behaviour for EEMI due to the current-induced dynamics of spin textures in $Gd_3Ru_4Al_{12}$ and $YMn_6Sn_6$. Moreover, we also observed the real part of the nonlinear complex impedance, partly resulting from the dissipative nature of the current-driven motion of helices – as well as domain walls – in these materials. These findings highlight the diverse current-nonlinear transport responses of spin dynamical origin in helimagnets.

**Acknowledgements**

The authors are grateful to R. Yamada for fruitful discussion. This work was supported by JSPS KAKENHI Grant Numbers 23H05431, 24K00566, and 24H01607 and Core Research for Evolutional Science and Technology (CREST), Japan Science and Technology Agency (JST) (Grant No. JPMJCR1874) as well as JST FOREST Grant No. JPMJFR2238 (Japan).

**Author contributions**

T.Y., Y. T. and N. N. conceived the project. T.Y. conducted transport measurements and analysed data for $Gd_3Ru_4Al_{12}$. A. K. and D. Y. conducted transport measurements and analysed data for $YMn_6Sn_6$. T.Y., M. H., N. N., and Y. T wrote the draft. All authors discussed the results and commented on the manuscript.